\documentclass[10pt,letterpaper,twocolumn]{article}

\usepackage{ol2}
\usepackage{amsmath}
\usepackage{amssymb}

\begin{document}

\twocolumn[

\title{Terahertz beat freuquency generation from two-mode lasing operation of coupled microdisk laser}

\author{Jung-Wan Ryu,$^1$ Jinhang Cho,$^2$ Chil-Min Kim,$^2$ Susumu Shinohara,$^3$ and Sang Wook Kim$^{4,*}$}
\address{$^1$Department of Physics, Pusan National University, Busan 609-735, Korea\\
$^2$Acceleration Research Center for Quantum Chaos Applications, Sogang University, Seoul 121-742, Korea\\
$^3$NTT Communication Science Laboratories, NTT Corporation, 2-4 Hikaridai, Seika-cho, Soraku-gun, Kyoto 619-0237, Japan\\
$^4$Department of Physics Education, Pusan National University, Busan 609-735, Korea\\
$^*$Corresponding author: swkim0412@pusan.ac.kr
}

\begin{abstract}
We propose a coupled microdisk laser as a compact and tunable laser source for the generation of a coherent continuous wave THz radiation by photomixing.
Using the Schr{\"o}dinger-Bloch model including the nonlinear effect of active medium,
we find single mode and two mode lasings depending on the pumping strength.
We explain the transitions of lasing modes in terms of resonant modes
which are the solutions of the Schr\"odinger-Bloch model without active medium and nonlinear interaction.
In particular, a two mode lasing is shown to generate THz oscillating frequency originating from the light beating of two nearly degenerated resonant modes
with different symmetries.
\end{abstract}

 ]

The Terahertz (THz) frequency range of the electromagnetic spectrum has attracted considerable research interests
due to its wide range of potential applications \cite{Ton07}.
Continuous-wave (CW) THz wave generation by photomixing, which enables the generation of CW THz
radiation by difference frequency generation of two lasers, was reported \cite{Bro93,Bro95,McI95}.
Because of the short photocarrier lifetime of a photomixer, the photocurrent is modulated at the difference frequency of two lasers,
so that the electromagnetic wave at this frequency is generated.
Recently, dual-mode lasers, which excite two laser modes from a single or combined laser cavity,
have been studied for the compact and tunable laser source for THz photomixing
in order to resolve the difficulty of alignment for spatial mode matching and stabilization of two independent lasers \cite{Hyo96,Tan00}.

Two-dimensional semiconductor microdisk lasers based on whispering-gallery mode (WGM)
have attracted much attention because of their high potential in a number of applications
related to high-density optoelectric integration \cite{MaC91,Yam93}.
The microdisk lasers can produce single lasing modes with ultralow threshold in dimensions of the order of an optical wavelength
due to strong light confinement by total internal reflection.
Advances in material science and nanofabrication techniques
enhance the utility of microdisk lasers, especially deformed \cite{Noe97,Gma98} and coupled \cite{Ben08,Ben11} microdisks.
Deformed and coupled microdisk lasers generate intrinsic mode
splittings of WGMs with degenerated clockwise and counterclockwise components \cite{Cha96,Ryu06,Bor07,Ryu09}.
The mode splittings of deformed and coupled microdisk lasers come from geometrical symmetry breakings due to the shape deformation and perturbation in the evanescent field outside
resonators, respectively.

In this Letter, we numerically study lasing operation of the coupled microdisk laser, employing the Schr\"odinger-Bloch (SB) model.
We demonstrate that the coupled microdisk laser can generate a tunable THz beatnote due to an intrinsic quasi-degeneracy of resonant modes.
Our results suggest that the coupled microdisk laser could be used as a compact laser source for the generation of THz radiation by photomixing.

Figure 1(a) shows two dimensional coupled microdisk laser.
We set the radii of right and left microdisks $R_{1}=1 \mu m$ and $R_{2}=0.95 \mu m$
and the interdisk distance $d=0.1 \mu m$.
The refractive indices inside and outside the two microdisks are $n_{in}=2.0$ and $n_{out}=1.0$.
In order to study the lasing modes in the coupled microdisk laser,
we use the SB model \cite{Har11,Har03a} considering the nonlinear interaction between light field and active medium
\begin{eqnarray}
\frac{\partial {\tilde E}}{\partial t} &=& \frac{i}{2} \left( \nabla^2_{xy}+\frac{n^2}{n^2_{in}} \right) {\tilde E}
- \alpha_{L}(x,y){\tilde E} + \mu {\tilde \rho}, \\
\frac{\partial {\tilde \rho}}{\partial t} &=& -\tilde \gamma_{\perp} \tilde \rho+\tilde \kappa W \tilde E, \\
\frac{\partial W}{\partial t} &=&
-\tilde \gamma_{\parallel}(W-W_{\infty})-2\tilde\kappa(\tilde E \tilde \rho^{*}+\tilde E^{*} \tilde \rho),
\end{eqnarray}
where $\tilde E$ and $\tilde \rho$ are the slowly varying envelope of the TM electric field and polarization field, respectively,
and $W$ is the population inversion.
The space and time are made dimensionless by the scale transformation
($n_{in}\omega_{0}x/c$, $n_{in}\omega_{0}y/c$, $t\omega_{0}$)$\rightarrow$($x$, $y$, $t$), respectively,
where $\omega_{0}$ is the oscillation frequency of the light field, which is the same as the transition frequency of two level medium
in our model.
The $\alpha_{L}(x,y)$ is the
linear absorption coefficient, which is the constant $\alpha_{L}$ inside the cavity and zero outside.
The transversal and longitudinal relaxation rates are made dimensionless
by the transformation ($\gamma_{\perp}/\omega_{0}$, $\gamma_{\parallel}/\omega_{0}$)$\rightarrow$
($\tilde \gamma_{\perp}$, $\tilde \gamma_{\parallel}$), respectively.
$\tilde \kappa$ is the dimensionless coupling strength $\kappa/\omega_{0}$.
$W_{\infty}$ is the external pumping parameter, which we control for transition of lasing modes.
Using this model, the nonlinear interaction between two (or more) lasing modes was studied in the stadium-shaped
microcavity laser \cite{Har03b,Shi05,Sun05,Har05}.

\begin{figure}[htb]
\centerline{\includegraphics[width=8.3cm]{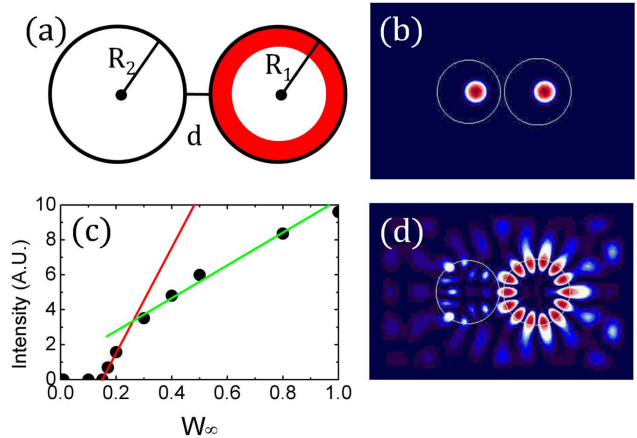}}
\caption{(Color online) (a) Schematic picture of coupled microdisk laser.
Pumping current is selectively injected to red shaded region in the SB model.
(b) The initial state of the light field in two microdisks.
(c) Total intensity of the light field inside two microdisks vs. pumping strength $W_\infty$.
(d) The light intensity pattern of final stable state at $W_\infty = 0.17$.
Red-white-blue-dark blue colors represent the intensity from high to low throughout this Letter.
}
\label{fig1}
\end{figure}

To improve lasing efficiency and to lase WGMs on the right microdisk,
we only pump near the outer boundary of the right microdisk.
The width of the selective pumping area in Fig.~1(a) is $0.3 R_{1}$.
The oscillation frequency $\omega_{0}$ of the light field, which lies at the gain center,
is $14.4185 \times 10^{14} \mathrm{Hz}$.
We fix the other parameters as follows: $\alpha_{L}=1.5074 \times 10^{-4}$, $\mu=26.7035$,
$\tilde \gamma_{\perp}=6.9400 \times 10^{-3}$, $\tilde \gamma_{\parallel}=6.9400 \times 10^{-6}$,
and $\tilde \kappa=1.1172 \times 10^{-5}$.

From the SB model, we obtain the time evolutions of intensities, the power spectrums, and the light intensity patterns of lasing modes as increasing
the pumping strength $W_{\infty}$.
The light field starts from an initial condition of two Gaussian wave packets, each being located inside a disk as shown in Fig.~1(b).
Figure~1(c) shows the total intensity of the light field inside two microdisks as a function of pumping strength $W_\infty$.
A single mode lasing occurs if the pumping power exceeds the lasing threshold, $W_{\infty}=0.16$.
In the regime of the single mode lasing, as time goes by, the total intensity of the light field grows exponentially at first
and finally becomes stationary through relaxation oscillation.
The optical spectrum of the lasing mode has a single peak and
its stationary intensity pattern is shown in Fig.~1(d).

In order to understand the lasing mode dynamics of the SB model,
it is useful to consider solutions, $\tilde E(x,y,t)=e^{-i \xi t} \psi(x,y)$ of a passive cavity obtained by putting $\alpha_L=\mu=0$ in Eq.~(1),
which are numerically solved by the boundary element method \cite{Har03a,Wie03}.
Figure~2 shows six high-Q resonant modes near the gain center, $\mathrm{Re}(\xi)=0$, and the intensity patterns of two
resonant modes closest to the gain center, with even (E) and odd (O) symmetries with respect to the $x$-axis.
It should be noted that the modes other than modes E and O in Fig.~2 are WGM localized on the left microdisk,
thus being unlikely to be excited in our selective pumping simulations.
The pattern of Fig.~1(d) is almost the same as the intensity pattern of the resonant mode E with even parity in Fig.~2.
These results convince us that the lasing mode is a single mode supported by resonant mode E, which is the highest-Q mode
near the gain center of the laser.

\begin{figure}[htb]
\centerline{\includegraphics[width=8.3cm]{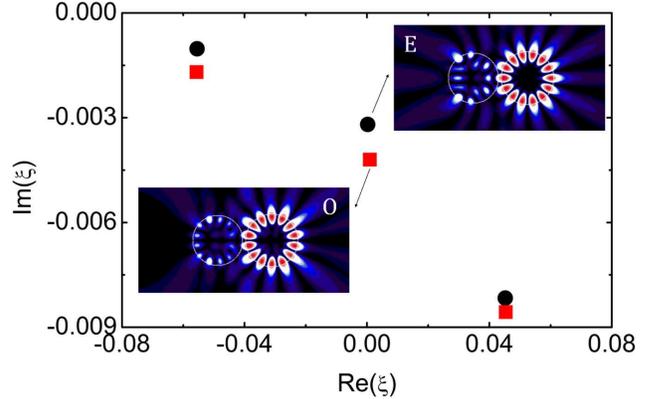}}
\caption{(Color online) Resonant modes with even (black circle) and odd (red square) symmetries near the gain center $\omega_0$
($\mathrm{Re}(\xi)=0$).
The intensity patterns E and O correspond to the resonant modes with even and odd symmetries, respectively.
}
\label{fig2}
\end{figure}

As we increase the pumping power, the single mode lasing changes into a two mode lasing when $W_{\infty}=0.25$.
Figure~1(c) shows the change of a slope of the light intensity at $W_\infty=0.25$, which represents a transition from single
mode lasing to two mode lasing. 
The time evolution of light intensity at $W_\infty=0.5$ is quasiperiodic as shown in Fig.~3(a), where the optical spectrum exhibits two peaks
corresponding to the modes E and O as shown in Fig.~2.
The quasiperiodic oscillation originates from the beating of the nearly degenerated modes E and O,
which is revealed by the fact that mode patterns at A, B, C, and D in Fig.~3(a) almost correspond to the linear superpositions of these two modes.

\begin{figure}[htb]
\centerline{\includegraphics[width=8.3cm]{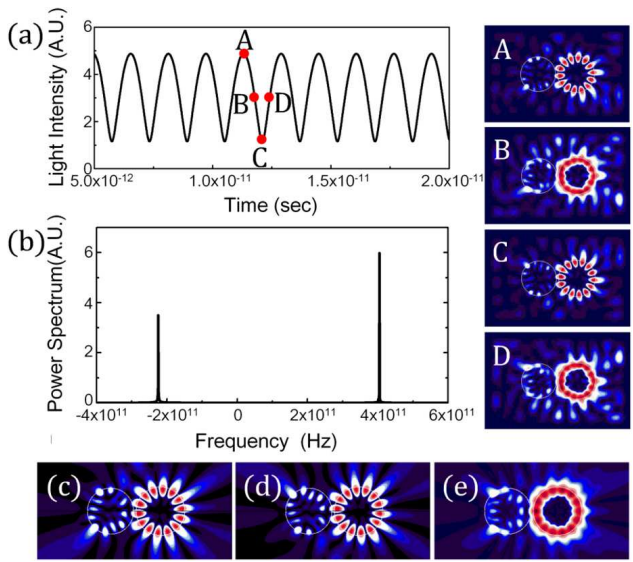}}
\caption{(Color online) (a) Time evolution of the intensity of light field on a point in space at $W_\infty=0.50$.
(b) The power spectrum obtained from the time evolution of the light field.
The patterns A, B, C, and D are the light intensity patterns of oscillating modes at times A, B, C, and D, respectively.
The intensity patterns of superposed modes
(c) $\left|\psi_{1}+\psi_{2}\right|^2$ and (d) $\left|\psi_{1}-\psi_{2}\right|^2$
of two modes E and O in Fig.~2.
The sum of intensities (e) $\left|\psi_{1}\right|^2+\left|\psi_{2}\right|^2$ of the two modes.
}
\label{fig3}
\end{figure}

When the interaction between modes E and O is small, the time evolution of the light field is
quasiperiodic as if there exist two independent modes of different frequencies $\nu_{1}$ and $\nu_{2}$.
The intensity varies sinusoidally at the beating frequency $\left|\nu_{1}-\nu_{2}\right|$,
0.6 THz in our case as confirmed in Fig.~3(b).
Since the time series of Fig.~3(a) is obtained at the observation point where there is no phase difference between modes E and O,
the maximum and the minimum of the intensity $I(t)$ are expressed as $\left|\psi_{1}+\psi_{2}\right|^2$ [Fig.~3(c)]
and $\left|\psi_{1}-\psi_{2}\right|^2$ [Fig.~3(d)],
respectively, and in between $\left|\psi_1\right|^2 + \left|\psi_2\right|^2$ [Fig.3(e)].
The oscillating intensity patterns are revealed to be the alternate oscillation in far field measurement,
which is a switching of light emission directions \cite{Cho08}.

Finally we emphasize the advantage of using the coupled microdisk laser in THz application.
First, it is relatively easy to achieve a two mode lasing in a coupled microdisk laser because
two high-Q modes with different symmetries are always very close to each other in mode frequency due to weak evanescent coupling.
Secondly the beat frequency of the two mode lasing in the coupled microdisk laser is tunable
because the mode splitting of two resonant modes can be controlled by changing various system parameters
such as the interdisk distance, the size of microdisk, and 
the refractive index which depends on the temperature of microdisk \cite{Ben11}.
For example, if the system is not too small, we can control systematically the beat frequency as a function of the interdisk
distance because the splitting of two nearly degenerate modes gets larger, i.e., the beat frequency increases as the interdisk distance decreases.
Thirdly the coupled microdisks have a benefit in optoelectric integration since they can generate a directional light emission
without influencing the spectral characteristics of the lasing modes.
This can be achieved simply by deforming the left microdisk where the light intensity is not strong \cite{Ryu11}.

In summary, we have studied the dynamics of lasing mode in coupled microdisk laser using the Schr\"odinger-Bloch model.
We have obtained single mode and two mode lasing regimes
as a function of pumping strength and explained the lasing mode dynamics based upon two modes of a passive cavity.
In the broad range of parameters,
the THz beat frequency can be generated by a quasiperiodic motion of the two mode lasing with nearly degenerate frequencies.
We expect a coupled microdisk laser to be a promising device as a compact and tunable laser source
for the generation of THz radiation by photomixing.

We would like to thank S.-Y. Lee and M. Choi for discussions.
This work was supported by the NRF grant funded by the Korea government (MEST) (No.2010-0024644 and No.2012R1A1A4A01013955).


\begin{thebibliography}{99}

\bibitem{Ton07} M. Tonouchi, Nat. Photon. {\bf 1}, 97 (2007).
\bibitem{Bro93} E.R. Brown, K.A. McIntosh, F.W. Smith, M.J. Manfra, and C.L. Dennis, Appl. Phys. Lett. {\bf 62}, 1206 (1993).
\bibitem{Bro95} E.R. Brown, K.A. McIntosh, K.B. Nichols, and C.L. Dennis, Appl. Phys. Lett. {\bf 66}, 285 (1995).
\bibitem{McI95} K.A. McIntosh, E.R. Brown, K.B. Nichols, O.B. McMahon, W.F. DiNatale, and T.M. Lyszczarz,
                Appl. Phys. Lett. {\bf 67}, 3844 (1995).
\bibitem{Hyo96} M. Hyodo, M. Tani, S. Matsuura, N. Onodera, and K. Sakai, Electron. Lett. {\bf 32}, 1589 (1996).                
\bibitem{Tan00} M. Tani, P. Gu, M. Hyodo, K. Sakai, and T. Hidaka, Opt. Quantum Electron. {\bf 32}, 503 (2000).                
\bibitem{MaC91} S.L. McCall, A.F.J. Levi, R.E. Slusher, S.J. Pearton, and R.A. Logan, Appl. Phys. Lett. {\bf 60}, 289 (1992).
\bibitem{Yam93} Y. Yamamoto and R.E. Slusher, Phys. Today {\bf 46} 66 (1993).
\bibitem{Noe97} J.U. N\"ockel and A.D. Stone, Nature (London) {\bf 385}, 45 (1997).
\bibitem{Gma98} C. Gmachl, F. Capasso, E.E. Narimanov, J.U.N\"ockel, A.D. Stone, J. Faist, D.L. Sivco, and A.Y. Cho,
                Science {\bf 280}, 1556 (1998).
\bibitem{Ben08} M. Benyoucef, S. Kiravittaya, Y.F. Mei, A. Rastelli, and O.G. Schmidt, Phys. Rev. B {\bf 77}, 035108 (2008).
\bibitem{Ben11} M. Benyoucef, J.-B. Shim, J. Wiersig, and O.G. Schmidt, Opt. Lett. {\bf 36}, 1317 (2011).
\bibitem{Cha96} {\it Optical Processes in Microcavities}, edited by R.K. Chang and A.K. Campillo (World Scientific, Singapore, 1996).
\bibitem{Ryu06} J.-W. Ryu, S.-Y. Lee, C.-M. Kim, and Y.-J. Park, Phys. Rev. A {\bf 74}, 013804 (2006).
\bibitem{Bor07} S.V. Boriskina, Opt. Lett. {\bf 32}, 1557 (2007).
\bibitem{Ryu09} J.-W. Ryu, S.-Y. Lee, and S. W. Kim, Phys. Rev. A 79, 053858 (2009).
\bibitem{Har11} T. Harayama and S. Shinohara, Laser and Photonics Reviews {\bf 5} 247 (2011).
\bibitem{Har03a} T. Harayama, P. Davis, and K.S. Ikeda, Phys. Rev. Lett. {\bf 90}, 063901 (2003).
\bibitem{Har03b} T. Harayama, T. Fukushima, S. Sunada, and K.S. Ikeda, Phys. Rev. Lett. {\bf 91}, 073903 (2003).
\bibitem{Shi05} S. Shinohara, S. Sunada, T. Harayama, and K.S. Ikeda, Phys. Rev. E {\bf 71}, 036203 (2005).
\bibitem{Sun05} S. Sunada, T. Harayama, and K.S. Ikeda, Phys. Rev. E {\bf 71}, 046209 (2005).
\bibitem{Har05} T. Harayama, S. Sunada, and K.S. Ikeda, Phys. Rev. A {\bf 72}, 013803 (2005).
\bibitem{Wie03} J. Wiersig, J. Opt. A: Pure Appl. Opt. {\bf 5} 53 (2003).
\bibitem{Cho08} M. Choi, T. Fukushima, and T. Harayama, Phys. Rev. A {\bf 77}, 063814 (2008).
\bibitem{Ryu11} J.-W. Ryu and M. Hentschel, Opt. Lett. {\bf 36} 1116 (2011).

\end{thebibliography}
\end{document}